# Answering Should Self-Publishing Video Game Developers Market Their Game on Twitter

Copenhagen Business School

**Nathaniel Francis Golding, Copenhagen Business School, Copenhagen, Denmark**
**Nago18ab@student.cbs.dk / nfgolding@gmail.com**

# Abstract

In the marketing of video games made by self-publishing independent developers, there is common advice to market a game using the social media platform Twitter. However, the advice that stems from industry sources is somewhat contradictory, and many self-publishing independent developers elect not to use Twitter at all. This presents an opportunity for researchers to investigate this tension and determine if using Twitter does have a causal influence on the successful release of a game by using relatively recent developments in causal data science techniques. In this sense, this paper highlights these causal inference developments while simultaneously informing self-publishing independent developers whether they should indeed market their games using Twitter. It was found that using Twitter results in an average increase of 85.4 reviews during release week, corresponding to a 314% positive difference. Using Twitter also doubles the chance of reaching a critical 10-review inflection point threshold on release week. These effects, however, are moderated by the characteristics of the game, expressed as 'tags', yet in no case is the effect of Twitter use reduced to 0. Based on these findings, it is advised that new, self-publishing video game developers do use Twitter to market their games.

# 1 Introduction

When it comes to marketing video games, there could be a large difference between Independent Developer's (Indie Devs) games that market on Twitter and those that do not. The premise here is that common advice in the industry instructs Indie Devs to use Twitter to market their game (zukalous, 2021b). Not to end-consumers per se, but influencers, journalists, and content aggregators. However, many games that get released by first-time self-publishing developers ignore this advice, with many games electing not to use Twitter to market their game before launch. Strangely enough, Zukalous also claims that having a "…[large] Twitter following is a result of a popular game, not the cause of it…" (zukalous, 2021b, 2021c). This raises the question, should developers market their games on Twitter, or not, since many games ignore the practice, and a revered industry expert sends mixed signals on the matter. It has been found in previous studies that there is a connection between engagement on Facebook, a Social Networking Site (SNS), and offline user behaviour (Mochon et al., 2017). This is great news, but Mochon et. al. (2017) focused on Facebook, the industry they focused on was health and wellness, for an existing customer base, and an existing brand. This establishes that it is possible that Twitter could affect purchasing behaviour, but does it in the case of video games? In this, admittedly simple study, we are interested in informing whether a new video game developer that self-publishes their game on Steam should use Twitter as a marketing medium. This query is also an excellent opportunity to use modern causal inference techniques, as relatively recent developments in this field have made it possible to understand when a causal effect can be estimated from purely observational data. To be clear, the work pioneered largely by Judea Pearl and Elias Barienboim (Bareinboim & Pearl, 2013a, 2013c, 2016; Bareinboim & Tian, 2015; Hünermund & Bareinboim, 2021; Pearl, 1995; Pearl et al., 2016; Pearl & Bareinboim, 2014, 2011; Pearl & Mackenzie, 2018) establishes that non-experimental[1] settings can be used to infer causal claims, as long as certain conditions are met. Namely if through the application of do-calculus, an appropriate identification strategy that eliminates do(X) - symbolising an experiment -, can be established (Pearl, 1995).

The problem formulation for this project is thus:

**What is the effect of marketing via Twitter on release-week reviews for new self-publishing video game developers?**

# 2 Theory Delimitation and Literature Review

In causal inference, the first rubric for establishing whether a causal effect can be discovered from observational data is to establish the Directed Acyclic Graph (DAG) for the causal query in question (Pearl & Mackenzie, 2018). This DAG is a conceptual framework that captures the relationships and assumptions of elements in a model, which then enables the principled application of rules and theorems that can judge if causal effects can be estimated from observable data, or if an experiment is indeed required (for a relatively concise overview of the state of the art see: Hünermund & Bareinboim, 2021; Pearl et al., 2016; Pearl & Mackenzie, 2018). In this chapter, we will consider industry sources that inform the conceptual framework and establish the DAG. Zukalous's write up about Twitter marketing (zukalous, 2021b) is the primary source of inspiration for developing our conceptual framework, as multiple industry connections referred specifically to his work as a reliable reference. Here, Zukalous states that Twitter use can build a pre-launch community for a game, and thus lead to more sales on launch week.

A game's sales data is difficult to acquire, however, but it has been established by Kontos (2021) that a game's reviews are highly correlated to a game's sales figures, which means we can use a game's reviews as a proxy for a game's sales. Zukalous (2021d) also writes about a minimum number of reviews that a game needs to have during its first week before one of Steam's algorithms will start to promote the game, with the exact number appearing to be 10 reviews. This echoes Styhre (2020, Chapter 6), who also reports that a video game needs to acquire a "critical mass" on launch to be successful. This establishes a time-horizon for our study, where we consider a cross-section of the game's release-week reviews and contrast developers that used Twitter before this point in time against those that did not. However, to establish that the Tweets were a cause of these release reviews, we need to consider confounders and their relationships with our outcome variable (release-week-reviews). Just like we are considering the SNS Twitter, and Mochon et. al (2017) considered Facebook, all other SNSs or in general, 'other marketing' also need to be factored into our DAG.

[1] Meaning there is no randomisation element

Other potential confounders are mediating (a thing that transfers an indirect or partial effect to our outcome variable) and moderating (a thing that strengthens or dilutes an effect on our outcome variable) effects of other elements. In our case, Zukalous (2020) has also identified variation in genre's wish lists - a form of pre-ordering or intention to purchase -, number of games released and median sales figures. This introduces the possibility of genres (a categorical proxy for game characteristics known on Steam more generally as "Tags") moderating the effect of Twitter on release-week reviews, as some genres are more popular than others. However, there are 448 different tags a game could have, and most games have several tags, usually up to 20 in total, so accounting for this is non-trivial. To this end, we could model it as a latent (unobserved) variable and attempt to compute the identification strategy for the direct effect (i.e., apply the rules of do-calculus to identify if the total effect of Twitter can still be estimated from observations despite ignoring tags). However, if we still want to consider the moderating effect on tags it needs to be modelled as an observed variable. After a little exploration of our dataset (see Appendix A), we may not need to consider all 448 tags after all. The overall distribution of the tags (in our dataset) reveals that there are what could be considered 'major tags' and 'minor tags', and the distributions of these 'major tags' are fairly consistent when comparing a game's first, second, or third tag across both those games that used Twitter, and those that did not. This means, that instead of checking all 448 tags, to check for moderating effects, we decided to only include the major tags, namely: "Indie", "Action", "Casual", "Adventure", "Singleplayer", "Simulation", "Action-Adventure", "3D", "Strategy", "2D", "RPG", "Early Access", "Multiplayer", "Puzzle", "Shooter", "VR", and "Platformer".

With these elements in consideration, we can construct our initial DAG – I say initial, as constructing the DAG, testing conditional independencies, and computing identification is usually an iterative affair.

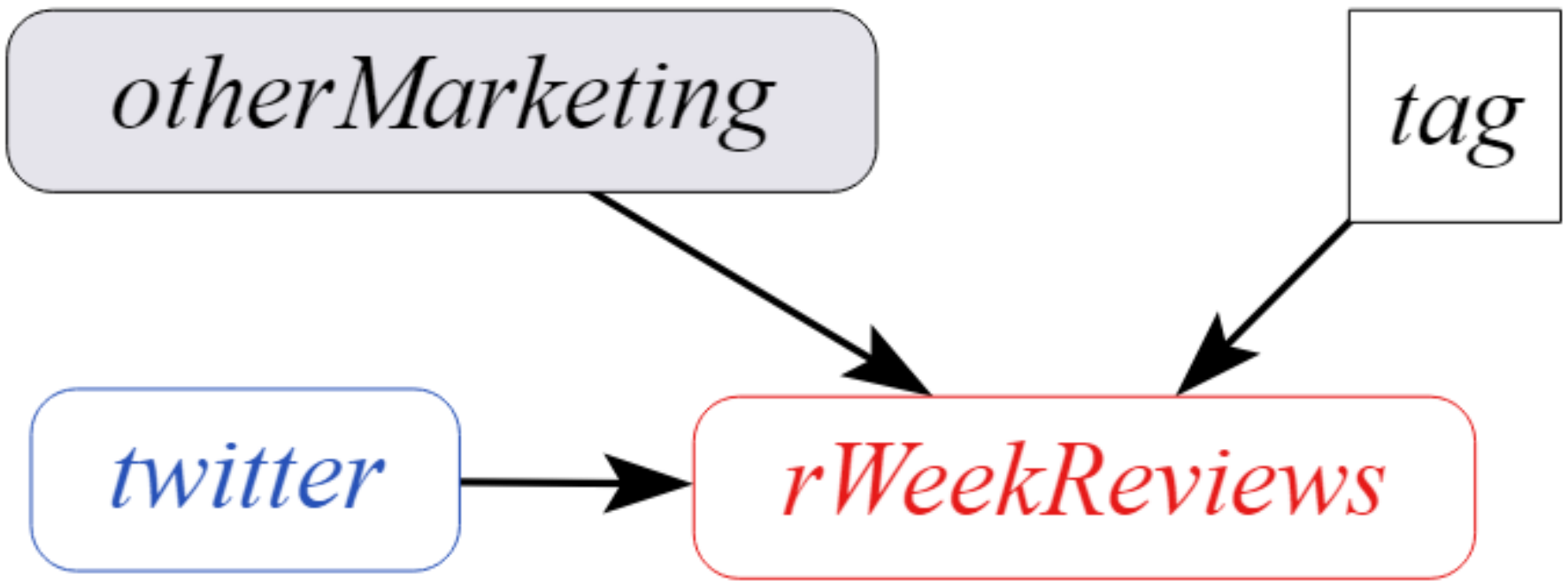

*Figure 1. Initial Directed Acyclic Graph for our query: causal impact of marketing via twitter on release-week reviews*

In Figure 1, the observed variable 'twitter' represents a binary dummy variable, - a value of 1 indicating that a game used Twitter in their pre-game-launch marketing, and 0 if not. This is our independent variable, or 'treatment' in the language of causal inference – i.e., 'do(x) is equivalent to 'do(treatment)'.

‘otherMarketing’ is a stand-in for every other marketing that a game could have used, in the initial DAG, we attempt to leave this as a latent variable (unobserved), as collecting information on each game’s marketing would also be a non-trivial task.
‘rWeekReviews’, is the number of reviews a game received during its first week after the official release, it is our dependent variable, or ‘outcome’ in the language of causal inference, the variable we are interested in estimating. One week was chosen based on the assumption that one week is long enough to cater for errors in Steam’s reporting of reviews, yet short enough to avoid significant effects of post-launch marketing efforts.
‘Tag’, refers to the potential moderator of a game’s characteristics, for simplification, it is represented in the diagram as a single node, but in the analysis, it will need to be expanded to a node for each tag. Where to include tag in the model is also open to debate. Here, we consider Pearl and Mackenzie’s (2018) concept of variables “listening” to each other to determine their state. Here, only the outcome variable ‘listens’ to tag, with Twitter and other-marketing’s state being independent of tag’s state (i.e., they do not ‘listen’ to tag to determine their own state).

As for computing the identification formula for our query ‘the causal effect of using Twitter on rWeekReviews’ the author used the web-based software casualfusion.net, which is being developed by Elias Barienboim and colleagues. Using causalfusion.net’s algorithmicised do-calculus, the causal query represented by the conceptual framework in the DAG, is indeed identifiable (Identification Formula 1). Meaning we may use purely observational data to estimate the causal effect, no experimental setting is required (see Figure 2 for a snippet of the identification formula computation in causalfusion.net).

$$P(rWeekReviews|do(twitter))=P(rWeekReviews|twitter)$$

*Identification Formula 1*

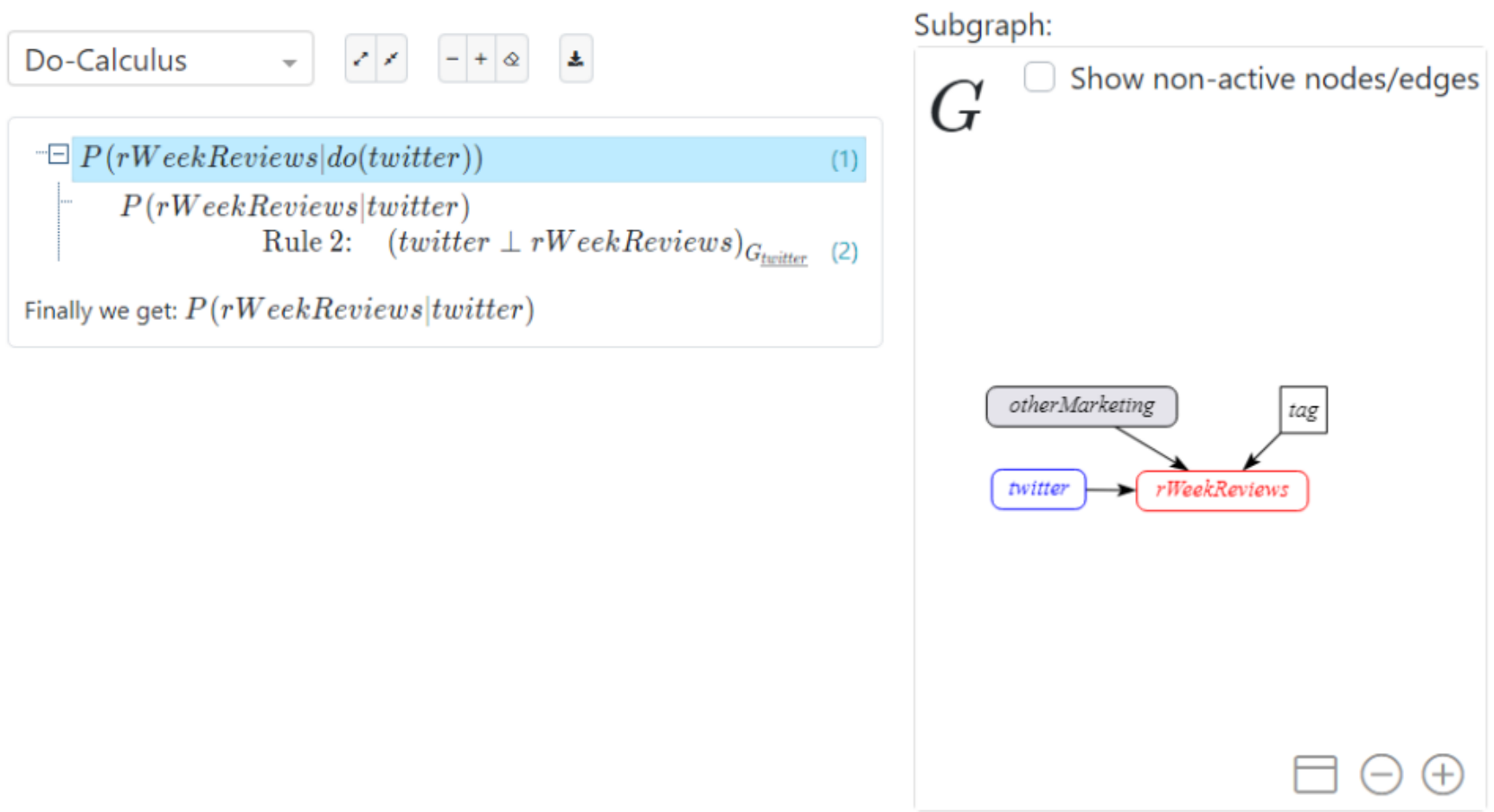

*Figure 2. Snippet of the application of do-calculus in causalfusion.net*

With the identification formula provided, we can now turn our attention to gathering the data and estimating the result. Note that in practice, data gathering, and computing identification formula is done simultaneously due to conditional independence tests for checking the DAG.

One major plus to this DAG in Figure 1 is that it is simple, however, the major downside to the framework is that it returned mixed results when attempting to test conditional independencies. According to d-separation rules (see Pearl et al., 2016, Chapter 2), it should be possible to check that twitter use is independent of tag ([tag]⫫twitter), and that tags should be independent of each other. We already know that tags are not independent of each other, as they describe characteristics of a game, which is not one-dimensional (for example, a game can be both 3D and adventure). Attempting to run a G Square conditional independence test for [tag]⫫twitter returned mixed results (see Appendix B). This mixed result could be an error in test selection, problems with the mixed data types (discrete numeric and binary) or indicate another confounding variable that isn't accounted for here. Thus, the reader should take this conceptual framework and its findings 'with a grain of salt'. Even though a DAG is a conceptual framework, to save on confusion, Figure 3 presents a more traditional conceptual framework for reference.

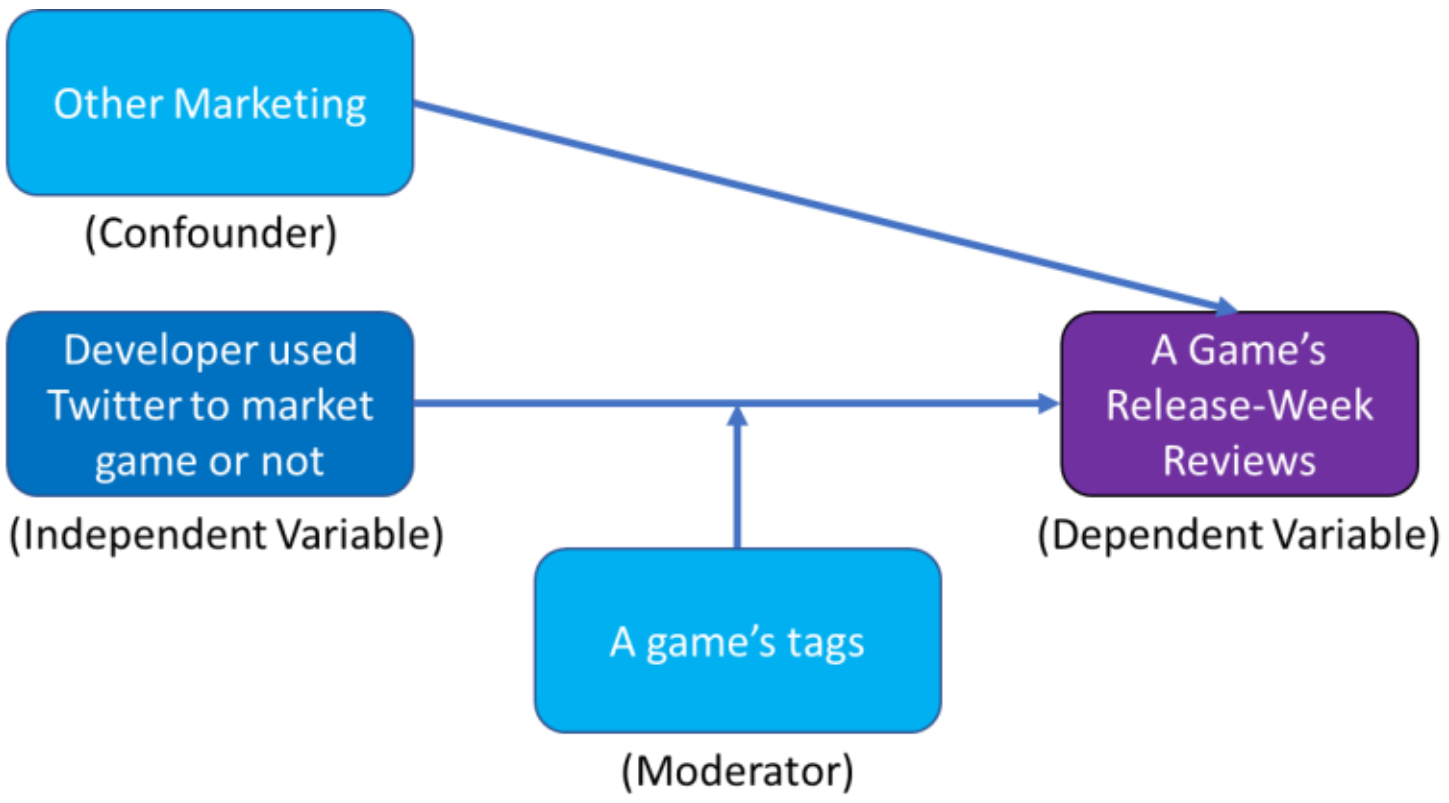

*Figure 3. Traditional conceptual framework representation of this research project*

# 3 Methodology

This chapter lays out the methodology used in gathering data and analysis. Data for a PC game released on steam was considered as the unit of analysis. Data were gathered as secondary data from Steam, SteamSpy.Com and GameDataCrunch.com. Secondary Data was chosen as it then ensures that the findings are based in 'real world data' which deflects criticism often laid on lab-based studies (Jung & Hinds, 2018).

## 3.1 Data Collection

A list of every game listed on steam from SteamSpy.com for every year up to 2020 (as 2021 data was still being populated, it was excluded from the dataset) was gathered (year-by-year) and compiled into a single list. This list from SteamSpy.com included the game, its release date, its developer, and its publisher. This list was then merged using python scripts with data generously provided by Lars Doucet from GameDataCrunch.com. Lar's data sets included the daily reviews for all games on Steam, each game's tags, and each game's listed SNS web address – if listed. The SNSs included Facebook, Twitch, YouTube, Twitter, Steam page and website.

Another python script was used to consolidate each game's daily reviews during its listed release week and merge this to the consolidated list of all games. At this stage, 40234 games were listed.

In this study, we are interested in informing new (meaning their first game) self-publishing developers however, so the list was filtered to exclude duplicate publishers and developers, leaving only their first title released – this reduced the population to 20260 games. This list was then further filtered to only include those developers that self-published their game, reducing the population to 17063. This list was then filtered again to only include games from 2013-2020, to account for the time it would take for Twitter to 'reach mainstream' based on Ross and Lemkin's (2016) assertion that it takes seven years for a company to no longer be considered a start-up, i.e., the company has reached 'mainstream' audiences (Moore, 2014). Thus, the final population was 16647. A dummy variable for Twitter was created and added to the dataset, plus another dummy variable for whether the game had more than 10 reviews based on Zukalous's (2021d) claim that 10 reviews are an inflection point – for use with logistic regression on a modified version of the DAG in

Figure 1. 1114 games used Twitter (568, 51%, of which had more than 10 reviews during release-week), and 15533 games that did not (3937, 25%, of which had more than 10 reviews during release-week). With these games' data gathered, we can turn to the analysis method.

## 3.2 Analysis methods

To analyse the data, two methods were used, both through the web-based platform causalfusion.net. Causalfusion.net was chosen, as the derived identification formula (see identification formula 1 in section 2) is non-parametric, and it was desirable for the software used to analyse the data to handle this, which causalfusion.net does. So, it made sense to continue using the same platform, though it is not strictly speaking necessary. Another benefit to using causalfusion.net is that it can combine multiple analysis models when estimating.

The first method used to analyse the data was a combination of XGBoost, Generalized Additive Models and Generalised Boosted Models. This method will give us the estimated (predicted) value of release-week-reviews (rWeekReviews), given a developer used Twitter (twitter == 1), or not (twitter == 0).
The second method used logistic regression, but swapped out rWeekReviews for moreThan10Reviews, a binary dependant variable that is equal to 1 if a game reached 10 reviews during the release week or 0 if it did not. Since the underlying DAG is unchanged (except for replacing the outcome variable), the identification formula derives to: P(moreThan10Reviews|do(twitter))=P(moreThan10Reviews|twitter); which is essentially the same as the previous formulation. The logistic regression will tell us the probability that a developer that marketed their game using Twitter would reach that 10-review inflection point.

Based on these two analyses, we will be able to understand the average effect that marketing via Twitter has on a game's release-week reviews (and by proxy, its sales). As for where and how to include tag's moderation in the analysis, this is open to debate. In an Ordinary Least Squares (OLS) linear regression, each tag would be included as an interaction, yet this might not be the most appropriate way in causal inference (see Rohrer et al., 2021's commentary). Instead, we will elect to split the data based on tag, then run the two analysis methods on each tag separately, reporting the effects in a table.

It is important to note here that we could have also elected to analyse the effect of the number of tweets, from a developer that tweets[2], on release-week reviews, yet deliberately chose not to do so in response to Rohrer et al.'s (2021) call that causal inference papers should be less ambitious in scope, more transparent and delivered in 'smaller pieces'. By delivering papers in this way the scientific community can more easily contribute to the affirmation or rejection of assumptions in a given causal model or conceptual framework, like our DAG. In leaving number of tweets on release-week reviews out, the community is given an opportunity to seriously consider and correct the assumptions we make, and thus iteratively build upon the framework presented. Each iteration of which better informs self-publishing video game developers on social media marketing imperatives.

[2] This would also incur a selection bias case which would need to be handled appropriately (see Bareinboim & Tian, 2015; Hünermund & Bareinboim, 2021)

# 4 Findings

This chapter presents the findings from the analysis methods described in section 3. It starts with presenting some descriptive statistics, then presents the findings from the two analyses in turn.

## 4.1 Descriptive Statistics

| *rWeekReviews* | *twitter == 0* | *twitter == 1* |
|---|---|---|
| Mean | 39.90059873 | 125.1229803 |
| Standard Error | 3.997296738 | 24.35290407 |
| Median | 3 | 11 |
| Mode | 0 | 0 |
| Standard Deviation | 498.1889153 | 812.8180766 |
| Sample Variance | 248192.1954 | 660673.2257 |
| Kurtosis | 6799.405088 | 789.8004851 |
| Skewness | 72.37086626 | 26.17860688 |
| Range | 50143 | 24991 |
| Minimum | 0 | 0 |
| Maximum | 50143 | 24991 |
| Sum | 619776 | 139387 |
| Count | 15533 | 1114 |
| Confidence Level (95.0%) | 7.835168214 | 47.78277672 |

*Table 1. Descriptive statistics for release-week reviews, comparing Twitter being used to not being used*

Table 1 shows the descriptive statistics for rWeekReviews divided by Twitter use. Here, we can see both the mean and median of twitter being used (twitter == 1) appears much higher than when it is not (twitter == 0). However, of particular note is the standard deviation and sample variance numbers. These are massive and indicates that our dataset is very ‘spread out’ and signals there could be large differences in the effectiveness of marketing campaigns, both when using Twitter and when not. Something that needs to be kept in mind as we move forward.

## 4.2 Twitter on rWeekReviews

| Experimental Distribution | | | | | | | | | | | | | | | | |
|---|---|---|---|---|---|---|---|---|---|---|---|---|---|---|---|---|
| | total (all tags) | | Indie | | Action | | Casual | | Adventure | | Singleplayer | | Simulation | | Action-Adventure | |
| *twitter* | 0 | 1 | 0 | 1 | 0 | 1 | 0 | 1 | 0 | 1 | 0 | 1 | 0 | 1 | 0 | 1 |
| **E[*rWeekReviews* \| do(*twitter*)]** | **39.9** | **125.3** | **35.0** | **129.1** | **49.0** | **160.1** | **29.4** | **85.9** | **39.9** | **125.3** | **72.1** | **187.4** | **78.7** | **192.3** | **55.9** | **250.4** |
| **σ(*rWeekReviews* \| do(*twitter*))** | 0.0 | 0.0 | 0.0 | 0.0 | 0.0 | 0.0 | 0.0 | 0.0 | 0.0 | 0.0 | 0.0 | 0.0 | 0.0 | 0.0 | 0.0 | 0.0 |
| **Count** | 8271 | 8376 | 6733 | 6736 | 4224 | 4258 | 3528 | 3476 | 8344 | 8303 | 2683 | 2695 | 1922 | 2006 | 1845 | 1775 |
| **Min** | 39.9 | 125.3 | 35.0 | 129.1 | 49.0 | 160.1 | 29.4 | 85.9 | 39.9 | 125.3 | 72.1 | 187.4 | 78.7 | 192.3 | 55.9 | 250.4 |
| **25%** | 39.9 | 125.3 | 35.0 | 129.1 | 49.0 | 160.1 | 29.4 | 85.9 | 39.9 | 125.3 | 72.1 | 187.4 | 78.7 | 192.3 | 55.9 | 250.4 |
| **Median** | 39.9 | 125.3 | 35.0 | 129.1 | 49.0 | 160.1 | 29.4 | 85.9 | 39.9 | 125.3 | 72.1 | 187.4 | 78.7 | 192.3 | 55.9 | 250.4 |
| **75%** | 39.9 | 125.3 | 35.0 | 129.1 | 49.0 | 160.1 | 29.4 | 85.9 | 39.9 | 125.3 | 72.1 | 187.4 | 78.7 | 192.3 | 55.9 | 250.4 |
| **Max** | 39.9 | 125.3 | 35.0 | 129.1 | 49.0 | 160.1 | 29.4 | 85.9 | 39.9 | 125.3 | 72.1 | 187.4 | 78.7 | 192.3 | 55.9 | 250.4 |
| Observational Distribution | | | | | | | | | | | | | | | | |
| **E[*rWeekReviews* \| *twitter*]** | 39.9 | 125.3 | 35.0 | 129.1 | 49.0 | 160.1 | 29.4 | 85.9 | 39.9 | 125.3 | 72.1 | 187.4 | 78.7 | 192.3 | 55.9 | 250.4 |
| **σ(*rWeekReviews* \| *twitter*)** | 498.1 | 813.9 | 249.1 | 884.6 | 668.7 | 1095.2 | 215.7 | 340.9 | 498.1 | 813.9 | 366.4 | 1069.6 | 906.1 | 466.2 | 457.8 | 1556.6 |
| **Count** | **15536** | **1111** | **12552** | **917** | **7910** | **572** | **6498** | **506** | **15536** | **1111** | **4750** | **628** | **3629** | **299** | **3346** | **274** |
| **Min** | 0 | 0 | 0 | 0 | 0 | 0 | 0 | 0 | 0 | 0 | 0 | 0 | 0 | 0 | 0 | 0 |
| **25%** | 1 | 3 | 1 | 3 | 1 | 3 | 0 | 2 | 1 | 3 | 1 | 5 | 1 | 5 | 1 | 4 |
| **Median** | 3 | 11 | 3 | 11 | 3 | 10 | 3 | 7 | 3 | 11 | 4 | 18 | 5 | 24 | 4 | 16 |
| **75%** | 11 | 44 | 11 | 43 | 11 | 48.5 | 9 | 32 | 11 | 44 | 18 | 76.25 | 17 | 119 | 13 | 72 |
| **Max** | 50143 | 24991 | 9333 | 24991 | 50143 | 24991 | 7897 | 4696 | 50143 | 24991 | 9333 | 24991 | 50143 | 4696 | 20362 | 24991 |
| **Parameters:** | | | | | | | | | | | | | | | | |
| **Prediction Algorithms: XGBoost, Generalized Additive Models, Generalized Boosted Models** | | | | | | | | | | | | | | | | |
| **p-value: 0.05** | | | | | | | | | | | | | | | | |
| **difference** | - | **85.4** | - | **94.1** | - | **111.1** | - | **56.6** | - | **85.4** | - | **115.3** | - | **113.6** | - | **194.4** |
| **difference %** | - | **314%** | - | **369%** | - | **327%** | - | **293%** | - | **314%** | - | **260%** | - | **244%** | - | **448%** |

| Experimental Distribution | | | | | | | | | | | | | | | | | | | | |
|---|---|---|---|---|---|---|---|---|---|---|---|---|---|---|---|---|---|---|---|---|
| | 3D | | Strategy | | 2D | | RPG | | Early Access | | Multiplayer | | Puzzle | | Shooter | | VR | | Platformer | |
| *twitter* | 0 | 1 | 0 | 1 | 0 | 1 | 0 | 1 | 0 | 1 | 0 | 1 | 0 | 1 | 0 | 1 | 0 | 1 | 0 | 1 |
| **E[*rWeekReviews* \| do(*twitter*)]** | **20.6** | **60.2** | **64.1** | **130.6** | **48.5** | **102.1** | **67.4** | **287.9** | **67.7** | **266.7** | **136.4** | **294.7** | **36.3** | **104.3** | **94.1** | **131.7** | **13.2** | **64.3** | **28.8** | **93.5** |
| **σ(*rWeekReviews* \| do(*twitter*))** | 0.0 | 0.0 | 0.0 | 0.0 | 0.0 | 0.0 | 0.0 | 0.0 | 0.0 | 0.0 | 0.0 | 0.0 | 0.0 | 0.0 | 0.0 | 0.0 | 0.0 | 0.0 | 0.0 | 0.0 |
| **Count** | 1685 | 1736 | 1712 | 1688 | 1652 | 1634 | 1585 | 1590 | 1572 | 1568 | 1358 | 1334 | 1262 | 1241 | 1127 | 1240 | 1035 | 1103 | 1036 | 1098 |
| **Min** | 20.6 | 60.2 | 64.1 | 130.6 | 48.5 | 102.1 | 67.4 | 287.9 | 67.7 | 266.7 | 136.4 | 294.7 | 36.3 | 104.3 | 94.1 | 131.7 | 13.2 | 64.3 | 28.8 | 93.5 |
| **25%** | 20.6 | 60.2 | 64.1 | 130.6 | 48.5 | 102.1 | 67.4 | 287.9 | 67.7 | 266.7 | 136.4 | 294.7 | 36.3 | 104.3 | 94.1 | 131.7 | 13.2 | 64.3 | 28.8 | 93.5 |
| **Median** | 20.6 | 60.2 | 64.1 | 130.6 | 48.5 | 102.1 | 67.4 | 287.9 | 67.7 | 266.7 | 136.4 | 294.7 | 36.3 | 104.3 | 94.1 | 131.7 | 13.2 | 64.3 | 28.8 | 93.5 |
| **75%** | 20.6 | 60.2 | 64.1 | 130.6 | 48.5 | 102.1 | 67.4 | 287.9 | 67.7 | 266.7 | 136.4 | 294.7 | 36.3 | 104.3 | 94.1 | 131.7 | 13.2 | 64.3 | 28.8 | 93.5 |
| **Max** | 20.6 | 60.2 | 64.1 | 130.6 | 48.5 | 102.1 | 67.4 | 287.9 | 67.7 | 266.7 | 136.4 | 294.7 | 36.3 | 104.3 | 94.1 | 131.7 | 13.2 | 64.3 | 28.8 | 93.5 |
| Observational Distribution | | | | | | | | | | | | | | | | | | | | |
| **E[*rWeekReviews* \| *twitter*]** | 20.6 | 60.2 | 64.1 | 130.6 | 48.5 | 102.1 | 67.4 | 287.9 | 67.7 | 266.7 | 136.4 | 294.7 | 36.3 | 104.3 | 94.1 | 131.7 | 13.2 | 64.3 | 28.8 | 93.5 |
| **σ(*rWeekReviews* \| *twitter*)** | 133.7 | 156.8 | 935.4 | 333.7 | 305.0 | 334.1 | 513.6 | 1823.1 | 965.1 | 2001.7 | 1198.6 | 1601.3 | 221.5 | 364.6 | 1213.4 | 415.6 | 62.4 | 149.4 | 210.2 | 238.3 |
| **Count** | **3224** | **197** | **3133** | **267** | **2895** | **391** | **2978** | **197** | **2982** | **158** | **2429** | **263** | **2227** | **276** | **2170** | **197** | **2043** | **95** | **1968** | **166** |
| **Min** | 0 | 0 | 0 | 0 | 0 | 0 | 0 | 0 | 0 | 0 | 0 | 0 | 0 | 0 | 0 | 0 | 0 | 0 | 0 | 0 |
| **25%** | 1 | 2 | 1 | 3 | 1 | 3 | 1 | 5 | 1 | 2 | 2 | 4 | 1 | 3 | 1 | 2 | 1 | 3 | 1 | 3 |
| **Median** | 3 | 8 | 4 | 16 | 4 | 13 | 5 | 30 | 4 | 9 | 8 | 27 | 4 | 10 | 4 | 9 | 3 | 7 | 3 | 10 |
| **75%** | 9 | 30 | 14 | 79.5 | 13 | 52 | 18 | 99 | 15 | 61.25 | 40 | 147 | 11 | 32.3 | 15 | 50 | 8 | 30.5 | 9 | 40 |
| **Max** | 5462 | 1283 | 50143 | 2329 | 7897 | 4696 | 20362 | 24991 | 50143 | 24991 | 50143 | 24991 | 5566 | 3688 | 50143 | 3688 | 1546 | 935 | 7059 | 1379 |
| **Parameters:** | | | | | | | | | | | | | | | | | | | | |
| **Prediction Algorithms: XGBoost, Generalized Additive Models, Generalized Boosted Models** | | | | | | | | | | | | | | | | | | | | |
| **p-value: 0.05** | | | | | | | | | | | | | | | | | | | | |
| **difference** | - | **39.6** | - | **66.5** | - | **53.6** | - | **220.5** | - | **199.0** | - | **158.3** | - | **67.9** | - | **37.5** | - | **51.1** | - | **64.6** |
| **difference %** | - | **292%** | - | **204%** | - | **211%** | - | **427%** | - | **394%** | - | **216%** | - | **287%** | - | **140%** | - | **488%** | - | **324%** |

*Table 2. Results of Twitter on rWeekReviews, including by tag*

Table 2 shows the results from running the causal query P(rWeekReviews|do(twitter))=P(rWeekReviews|twitter) in causalfusion.net using the combination of the XGBoost, Generalised Additive Models and Generalised Boosted Models prediction algorithms. In the case of the direct effect, Twitter use causes an average of an additional 85.4 release-week reviews, corresponding to a positive 314% difference on average to release-week-reviews. Where the moderating effects of tags were considered, every tag still saw an increase in release-week reviews when Twitter was used, but the amount did vary between tags. Tag "VR" saw the biggest difference in reviews (488%), and tag "Shooter" saw the smallest (140%). From these results, we can conclude two things. First, using Twitter to market a game, before the release of said game, leads to more reviews on average for a game. Second, a game's tag(s), does have a moderating effect on a game's release-week reviews, although in no case did the moderating effect reduce the effect of using Twitter to 0. All results are reported with a p-value of at least 0.05, this was set as a condition when executing the algorithms on causalfusion.net. However, the actual p-value of the findings is not reported, this could be an oversight from the developer, yet it means that these findings may have a higher significance which is not reported. A k-fold cross-validation method is used by causalfusion.net to validate the estimation, 3 folds were used, yet again the accuracy of the estimate folding is not reported. It is also worth noticing that the standard deviation in the observational distribution is quite large, in all cases. Meaning that even though there is an improved mean from using Twitter, the actual effect coming from the observations appears to vary wildly and is quite concerning.

## 4.3 Twitter on moreThan10Reviews

Table 3 shows the results of running the logistic regression on a developer using Twitter and a game achieving more than 10 reviews on release,

| Experimental Distribution | total (all tags) | | Indie | | Action | | Casual | | Adventure | | Singleplayer | |
|---|---|---|---|---|---|---|---|---|---|---|---|---|
| *twitter* | 0 | 1 | 0 | 1 | 0 | 1 | 0 | 1 | 0 | 1 | 0 | 1 |
| **E[*moreThan10Reviews* \| do(*twitter*)]** | **0.25** | **0.51** | **0.26** | **0.50** | **0.26** | **0.49** | **0.22** | **0.44** | **0.25** | **0.51** | **0.34** | **0.60** |
| **σ(*moreThan10Reviews* \| do(*twitter*))** | 0.0 | 0.0 | 0.0 | 0.0 | 0.0 | 0.0 | 0.0 | 0.0 | 0.0 | 0.0 | 0.0 | 0.0 |
| **Count** | 8341 | 8306 | 6709 | 6760 | 4255 | 4227 | 3517 | 3487 | 8303 | 8344 | 2682 | 2696 |
| **Min** | 0.25 | 0.51 | 0.26 | 0.50 | 0.3 | 0.5 | 0.2 | 0.4 | 0.3 | 0.5 | 0.3 | 0.6 |
| **25%** | 0.25 | 0.51 | 0.26 | 0.50 | 0.3 | 0.5 | 0.2 | 0.4 | 0.3 | 0.5 | 0.3 | 0.6 |
| **Median** | 0.25 | 0.51 | 0.26 | 0.50 | 0.3 | 0.5 | 0.2 | 0.4 | 0.3 | 0.5 | 0.3 | 0.6 |
| **75%** | 0.25 | 0.51 | 0.26 | 0.50 | 0.3 | 0.5 | 0.2 | 0.4 | 0.3 | 0.5 | 0.3 | 0.6 |
| **Max** | 0.25 | 0.51 | 0.26 | 0.50 | 0.3 | 0.5 | 0.2 | 0.4 | 0.3 | 0.5 | 0.3 | 0.6 |
| Observational Distribution | | | | | | | | | | | | |
| **E[*moreThan10Reviews* \| *twitter*]** | 0.25 | 0.51 | 0.26 | 0.50 | 0.26 | 0.49 | 0.22 | 0.44 | 0.25 | 0.51 | 0.34 | 0.60 |
| **σ(*moreThan10Reviews* \| *twitter*)** | 0.44 | 0.50 | 0.44 | 0.50 | 0.44 | 0.50 | 0.41 | 0.50 | 0.44 | 0.50 | 0.47 | 0.49 |
| **Count** | 15536 | 1111 | 12552 | 917 | 7910 | 572 | 6498 | 506 | 15536 | 1111 | 4750 | 628 |
| **Min** | 0 | 0 | 0 | 0 | 0 | 0 | 0 | 0 | 0 | 0 | 0 | 0 |
| **25%** | 0 | 0 | 0 | 0 | 0 | 0 | 0 | 0 | 0 | 0 | 0 | 0 |
| **Median** | 0 | 1 | 0 | 1 | 0 | 0 | 0 | 0 | 0 | 1 | 0 | 1 |
| **75%** | 1 | 1 | 1 | 1 | 1 | 1 | 0 | 1 | 1 | 1 | 1 | 1 |
| **Max** | 1 | 1 | 1 | 1 | 1 | 1 | 1 | 1 | 1 | 1 | 1 | 1 |
| **Parameters** | | | | | | | | | | | | |
| **Prediction Algorithms: Logistic Regression** | | | | | | | | | | | | |
| **Number of folds: 3** | | | | | | | | | | | | |
| **p-value: 0.05** | | | | | | | | | | | | |
| **Difference** | - | **0.26** | - | **0.25** | - | **0.23** | - | **0.22** | - | **0.26** | - | **0.27** |
| **Difference as %** | - | **201%** | - | **195%** | - | **191%** | - | **202%** | - | **201%** | - | **180%** |

| Experimental Distribution | Simulation | | Action-Adventure | | 3D | | Strategy | | 2D | | RPG | |
|---|---|---|---|---|---|---|---|---|---|---|---|---|
| *twitter* | 0 | 1 | 0 | 1 | 0 | 1 | 0 | 1 | 0 | 1 | 0 | 1 |
| **E[*moreThan10Reviews* \| do(*twitter*)]** | **0.33** | **0.65** | **0.29** | **0.60** | **0.21** | **0.45** | **0.30** | **0.57** | **0.28** | **0.54** | **0.34** | **0.64** |
| **σ(*moreThan10Reviews* \| do(*twitter*))** | 0.0 | 0.0 | 0.0 | 0.0 | 0.0 | 0.0 | 0.0 | 0.0 | 0.0 | 0.0 | 0.0 | 0.0 |
| **Count** | 2007 | 1921 | 1758 | 1862 | 1725 | 1696 | 1662 | 1738 | 1626 | 1660 | 1592 | 1583 |
| **Min** | 0.3 | 0.6 | 0.3 | 0.6 | 0.2 | 0.4 | 0.3 | 0.6 | 0.3 | 0.5 | 0.3 | 0.6 |
| **25%** | 0.3 | 0.6 | 0.3 | 0.6 | 0.2 | 0.4 | 0.3 | 0.6 | 0.3 | 0.5 | 0.3 | 0.6 |
| **Median** | 0.3 | 0.6 | 0.3 | 0.6 | 0.2 | 0.4 | 0.3 | 0.6 | 0.3 | 0.5 | 0.3 | 0.6 |
| **75%** | 0.3 | 0.6 | 0.3 | 0.6 | 0.2 | 0.4 | 0.3 | 0.6 | 0.3 | 0.5 | 0.3 | 0.6 |
| **Max** | 0.3 | 0.6 | 0.3 | 0.6 | 0.2 | 0.4 | 0.3 | 0.6 | 0.3 | 0.5 | 0.3 | 0.6 |
| Observational Distribution | | | | | | | | | | | | |
| **E[*moreThan10Reviews* \| *twitter*]** | 0.33 | 0.65 | 0.29 | 0.60 | 0.21 | 0.45 | 0.30 | 0.57 | 0.28 | 0.54 | 0.34 | 0.64 |
| **σ(*moreThan10Reviews* \| *twitter*)** | 0.47 | 0.48 | 0.45 | 0.49 | 0.41 | 0.50 | 0.46 | 0.50 | 0.45 | 0.50 | 0.47 | 0.48 |
| **Count** | 3629 | 299 | 3346 | 274 | 3224 | 197 | 3133 | 267 | 2895 | 391 | 2978 | 197 |
| **Min** | 0 | 0 | 0 | 0 | 0 | 0 | 0 | 0 | 0 | 0 | 0 | 0 |
| **25%** | 0 | 0 | 0 | 0 | 0 | 0 | 0 | 0 | 0 | 0 | 0 | 0 |
| **Median** | 0 | 1 | 0 | 1 | 0 | 0 | 0 | 1 | 0 | 1 | 0 | 1 |
| **75%** | 1 | 1 | 1 | 1 | 0 | 1 | 1 | 1 | 1 | 1 | 1 | 1 |
| **Max** | 1 | 1 | 1 | 1 | 1 | 1 | 1 | 1 | 1 | 1 | 1 | 1 |
| **Parameters** | | | | | | | | | | | | |
| **Prediction Algorithms: Logistic Regression** | | | | | | | | | | | | |
| **Number of folds: 3** | | | | | | | | | | | | |
| **p-value: 0.05** | | | | | | | | | | | | |
| **Difference** | - | **0.31** | - | **0.31** | - | **0.23** | - | **0.26** | - | **0.26** | - | **0.30** |
| **Difference as %** | - | **194%** | - | **209%** | - | **209%** | - | **186%** | - | **193%** | - | **190%** |

| Experimental Distribution | Early Access | | Multiplayer | | Puzzle | | Shooter | | VR | | Platformer | |
|---|---|---|---|---|---|---|---|---|---|---|---|---|
| *twitter* | 0 | 1 | 0 | 1 | 0 | 1 | 0 | 1 | 0 | 1 | 0 | 1 |
| **E[*moreThan10Reviews* \| do(*twitter*)]** | **0.32** | **0.47** | **0.45** | **0.61** | **0.25** | **0.50** | **0.32** | **0.44** | **0.20** | **0.45** | **0.23** | **0.48** |
| **σ(*moreThan10Reviews* \| do(*twitter*))** | 0.0 | 0.0 | 0.0 | 0.0 | 0.0 | 0.0 | 0.0 | 0.0 | 0.0 | 0.0 | 0.0 | 0.0 |
| **Count** | 1578 | 1562 | 1349 | 1343 | 1218 | 1285 | 1188 | 1179 | 1064 | 1074 | 1068 | 1066 |
| **Min** | 0.3 | 0.5 | 0.5 | 0.6 | 0.3 | 0.5 | 0.3 | 0.4 | 0.2 | 0.5 | 0.2 | 0.5 |
| **25%** | 0.3 | 0.5 | 0.5 | 0.6 | 0.3 | 0.5 | 0.3 | 0.4 | 0.2 | 0.5 | 0.2 | 0.5 |
| **Median** | 0.3 | 0.5 | 0.5 | 0.6 | 0.3 | 0.5 | 0.3 | 0.4 | 0.2 | 0.5 | 0.2 | 0.5 |
| **75%** | 0.3 | 0.5 | 0.5 | 0.6 | 0.3 | 0.5 | 0.3 | 0.4 | 0.2 | 0.5 | 0.2 | 0.5 |
| **Max** | 0.3 | 0.5 | 0.5 | 0.6 | 0.3 | 0.5 | 0.3 | 0.4 | 0.2 | 0.5 | 0.2 | 0.5 |
| Observational Distribution | | | | | | | | | | | | |
| **E[*moreThan10Reviews* \| *twitter*]** | 0.32 | 0.47 | 0.45 | 0.61 | 0.25 | 0.50 | 0.32 | 0.44 | 0.20 | 0.45 | 0.23 | 0.48 |
| **σ(*moreThan10Reviews* \| *twitter*)** | 0.46 | 0.50 | 0.50 | 0.49 | 0.43 | 0.50 | 0.47 | 0.50 | 0.40 | 0.50 | 0.42 | 0.50 |
| **Count** | 2982 | 158 | 2429 | 263 | 2227 | 276 | 2170 | 197 | 2043 | 95 | 1968 | 166 |
| **Min** | 0 | 0 | 0 | 0 | 0 | 0 | 0 | 0 | 0 | 0 | 0 | 0 |
| **25%** | 0 | 0 | 0 | 0 | 0 | 0 | 0 | 0 | 0 | 0 | 0 | 0 |
| **Median** | 0 | 0 | 0 | 1 | 0 | 0 | 0 | 0 | 0 | 0 | 0 | 0 |
| **75%** | 1 | 1 | 1 | 1 | 1 | 1 | 1 | 1 | 0 | 1 | 0 | 1 |
| **Max** | 1 | 1 | 1 | 1 | 1 | 1 | 1 | 1 | 1 | 1 | 1 | 1 |
| **Parameters** | | | | | | | | | | | | |
| **Prediction Algorithms: Logistic Regression** | | | | | | | | | | | | |
| **Number of folds: 3** | | | | | | | | | | | | |
| **p-value: 0.05** | | | | | | | | | | | | |
| **Difference** | - | **0.15** | - | **0.16** | - | **0.24** | - | **0.12** | - | **0.25** | - | **0.25** |
| **Difference as %** | - | **149%** | - | **135%** | - | **196%** | - | **139%** | - | **225%** | - | **208%** |

*Table 3. Results of Logistic Regression twitter on moreThan10Reviews, including by tag*

achieving the inflection point on release day that satisfies Styhre's (2020) critical mass, and Zukalous's (2021d) inflection point during release-week. In the average direct effects case (total, all tags), using Twitter corresponds to an increase of 0.26 in the probability that a game will achieve 10 reviews during release-week (an increase from 25% chance to 51% chance of achieving more than 10 reviews). This corresponds to a 201% difference. In the moderated cases where a tag is considered, we again see that the probability of achieving more than 10 reviews changes with each tag. The largest difference is found in the "VR" tag, with a 225% difference (in favour of using Twitter), and the smallest is for the "Multiplayer" tag, with a 135% difference (again, in favour of using Twitter). With these findings, also considering the findings from the Twitter on rWeekReviews section we can conclude that using Twitter to market a game before the release of that game, increases the probability of that game exceeding that critical 10 review inflection point. However, this also indicates that using Twitter only increases the probability, it does not guarantee more than 10 reviews will be achieved. Again, we can also conclude that a game's characteristics (i.e., its tagging) do influence that probability of success. As with Twitter on rWeekReviews, all results are reported with a p-value of at least 0.05, and a k-fold cross-validation method with 3 folds was used by causalfusion.net to validate the estimation.

# 5 Discussion

#### *5.1.1 Theoretical Implications*

From an academic perspective, the research carried out here is far too simplistic – likely owing to its departure point in industry-based knowledge, making it a case of 'applied science', rather than 'basic science' (Saunders et al., 2019). However, the findings here form a stable base from which other, more interesting research could be conducted. The findings from Twitter on rWeekReviews indicated that including a Twitter marketing campaign before releasing a game increases the average number of reviews. Yet, the findings from Twitter on moreThan10Reviews found that the increase in release-week reviews is not a given, it is only an increase in the probability of reaching that 10-review critical threshold that is provided by a Twitter campaign. When we view these two findings together, along with the significant variation displayed, we learn that just having a Twitter campaign is not enough, but it is a start. We need to think about sources of that variation, and what could cause changes to the probability of success. Here, we can turn to other theories within social media marketing for explanations. A likely contributor is the quality of a campaign itself, meaning the content of the posts, the copy, pictures, community-management dynamics, and such. That is to say, how a Twitter campaign is executed should also affect the effectiveness of using Twitter to market a game, not just whether a campaign is attempted. Li and Xie (2020) found that the mere presence of pictures affects user engagement, as does the characteristics of a picture.

This is similar to Zukalous's (2021a) assertations on a "beautiful/GIFable game" – that images and gifs of a gameplay an important part in attracting attention to that game. Berger and Milkman (2012) also argue for content effects, with their findings indicating that the emotional valence (positive emotions versus negative emotions) also affect the virality of posts, yet do so in connection to the arousal of a post (how well it triggers an emotional response from a user). They found that in general, positive posts are better than negative ones, yet high arousal trumps emotions. This is somewhat of a contradiction to Zukalous's (2021a) claim that Twitter will either love a game or not, which suggests a developer cannot influence reactions to posted content. Individual posts, however, need also to be considered in the context of an entire campaign, as Shehu et al., (2016) found that likeability dynamics affect evaluations of online advertisement. Viewing those findings in light of those here, suggests that the dynamics in a campaign over time are also important to consider. I.e., a campaign is made of many single posts, and controlling the likability of content over time is a contributor to a campaign's success.

It is also important to note that in this study, all tweet usage was considered as equal, yet Twitter does have a promotion function, which can increase the reach of a tweet to more people. This promotion feature could also be interesting to add as a variable in future research, to see if developers that used Twitter's promotion feature increase either release-week reviews or the probability of reaching more than 10 release-week reviews. This would make the research here more comparable to the research conducted by Mochon et al., (2017), who found that "Boosting" (the Facebook equivalent to Twitter's "Promoting") had a positive effect on customer behaviour, but organically acquired engagement did not. In contrast, the study here suggests all firm-initiated Twitter usage would lead to increased release-week reviews, but this is based on ignoring the potential of promoted tweets to change the outcome. However, the departure point for this research is based on common industry advice from Zukalous, and in the advice that he presents to developers, he makes no mention of using Twitter's promotion feature, thus we find it somewhat safe to assume that self-publishing developers would follow Zukalous's (2021a, 2021b) advice strictly, and thus not consider Twitter's promotion feature – although it is acknowledged this could be a factor that needs further investigation.

Concerning brand community theories, the research here focused on new developers self-publishing their first games, to this end, our findings might question self-identification idealisms like the idea of an "extended self" (Belk, 2013). That is, Mochon et al., (2017) found that simply inviting users to follow a page was enough to trigger user behaviour. In our study, by focusing on developers that have no existing communities for which a consumer could self-identify, we also consider that consumers engage with a new video game developer, simply because they could, not because of strong identity congruence. Similarly, Logan's (2014) findings suggested that consumers' decision to follow a brand online is more akin to "impulse buying" rather than planned behaviour. This may have relevance for the type of content that a developer posts to Twitter, either strong identity signalling content, or content that is more hedonistic, such as information or entertainment as per findings from Tsai and Men (2013). Although this may simply be a chicken-or-the-egg problem, where-in one type of content essentially leads to determining the other, it is another possible dynamic to consider in future research.

With the findings of this study as a baseline, we can in future studies introduce content effect measures and or self-identification concepts, including measures over time, and learn how these other theories change the effectiveness of a Twitter campaign. In effect, guiding developers of a game towards more effective campaigns, and thus more reviews for their games. Keeping in mind that reviews are a proxy for sales, more reviews also mean more sales.

#### *5.1.2 Managerial Implications*

Here we answer the managerial query that underlies our research question, essentially transforming the "what is the effect of Twitter use on release-week-reviews?" into a "should video game developers use Twitter to market their games, before release?". From our findings, it is advised that video game developers should indeed market their games on Twitter, as doing so doubles their chances of reaching the 10-review threshold (on average). However, self-publishing video game developers should also consider that success is not guaranteed, with success still being affected by factors not considered in this study. To that end, developers should look to other sources of theory, like those mentioned above, when designing their campaigns. The characteristics of a game are also something developers (and generally, marketing managers) need to consider, as tags do moderate the effect from Twitter. This means that Twitter users will respond differently depending on the game's type and nature. It is indeed probable that other specific SNSs suit specific types (tags) of games, but since this study is limited to only twitter use, it is not something that we can inform on here.

#### *5.1.3 Limitations and future research*

One significant limitation to the validity of the findings here is the mixed testable implications in the DAG. What this means is that we can't be sure that the findings here are true because we can't rule out alternative relationship arrangements in the model by falsification. However, if we consider this research forming a baseline for future research that considers additional parameters, such as content effects, then the testable implications are a factor that can be addressed in future DAGs. We know from the findings that simply doing a Twitter campaign is not enough, we also need to conduct future research that considers how to increase the effectiveness of a campaign. Future DAGs that build on the one used in this study will include more variables, thus providing more conditionally independent relationships to test on. This is exactly the idea behind keeping the DAG and analysis simple, in line with the call from Roher et al., (2021). Future research can critically evaluate the DAG in this paper, as they attempt to develop the framework further.

Another limitation of this research is the external validity of the findings. From the outset, this research paper aimed at informing developers in the video game industry, and thus the findings pertain only to the video game industry. It is possible that marketing new products on Twitter would be beneficial for other industries, but it is impossible to conclude transportability based on this finding alone. That being said, meta-transportability (Bareinboim & Pearl, 2013b) of these findings may still be possible, where the findings here are paired up with similar findings from other studies. Thus, this study also becomes useful for meta-analyses that could make more generalised claims. Where this studies' findings could transfer to, would be the case of non-self-publishing developers, and publishing beyond the first game. Although we specifically limited the case studied to first-time self-publishing developers, it is highly likely that releasing an additional game would be even more effective, as the established follower base would likely not revert to 0. Meaning the developer would not be starting from scratch for the additional campaign, instead, continuing to build a brand community that was established during the first Twitter campaign.

Future research should also look into the effects of other SNSs on release-week reviews in a similar fashion to the study done here. This is mostly to continue the theme of applied research, which attempts to bring established theories closer to the problems managers face, such as the strategic selection of SNSs for marketing games. Conducting a similar study on Reddit, Discord, YouTube, and Twitch will add to the body of literature on video game marketing, an area that is largely dominated by anecdotal evidence and opinion pieces.

# 6 Conclusion

In this study, we analysed the effect of using Twitter for marketing a video game on the number of release-week-reviews a game receives. This was done to inform new self-publishing video game developers on whether they should use Twitter, or not, since many first-time developers do not use Twitter, and an industry expert sends mixed signals on the use of Twitter (zukalous, 2021a, 2021b, 2021c). We used causal inference methods to compute an identification formula that enabled the use of observational data to estimate the effect. Our findings indicate that on average using Twitter increases the number of release-week-reviews by 85.4 reviews, corresponding to a positive 314% difference. However, we also found that using Twitter only increases the probability, roughly doubling the chance, of reaching the 10-review inflection point required (zukalous, 2021d). In both cases, the tag a game has does have a moderating effect. Based on these findings, it is advised that new, self-publishing video game developers do use Twitter to market their games yet look to other sources for exactly how to execute a Twitter campaign – such as content effects theories.

# 7 Acknowledgements

A big thank-you goes to Lars Doucet from gamedatacrunch.com, who provided several datasets that were merged to provide a population-level dataset.

# 9 Appendices:

## Appendix A: Visualisation of the count of each tag in the dataset

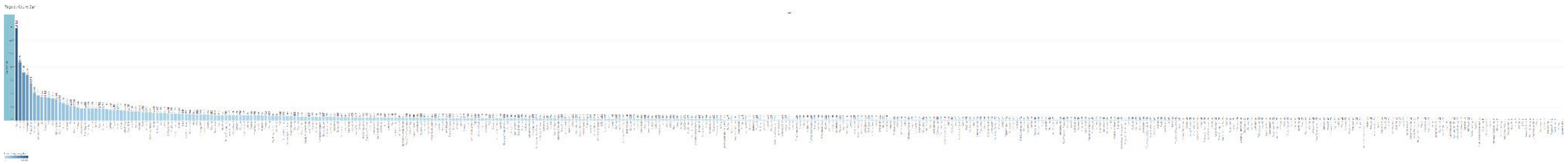

*Figure 4. Bar chart showing count of each tag in dataset*

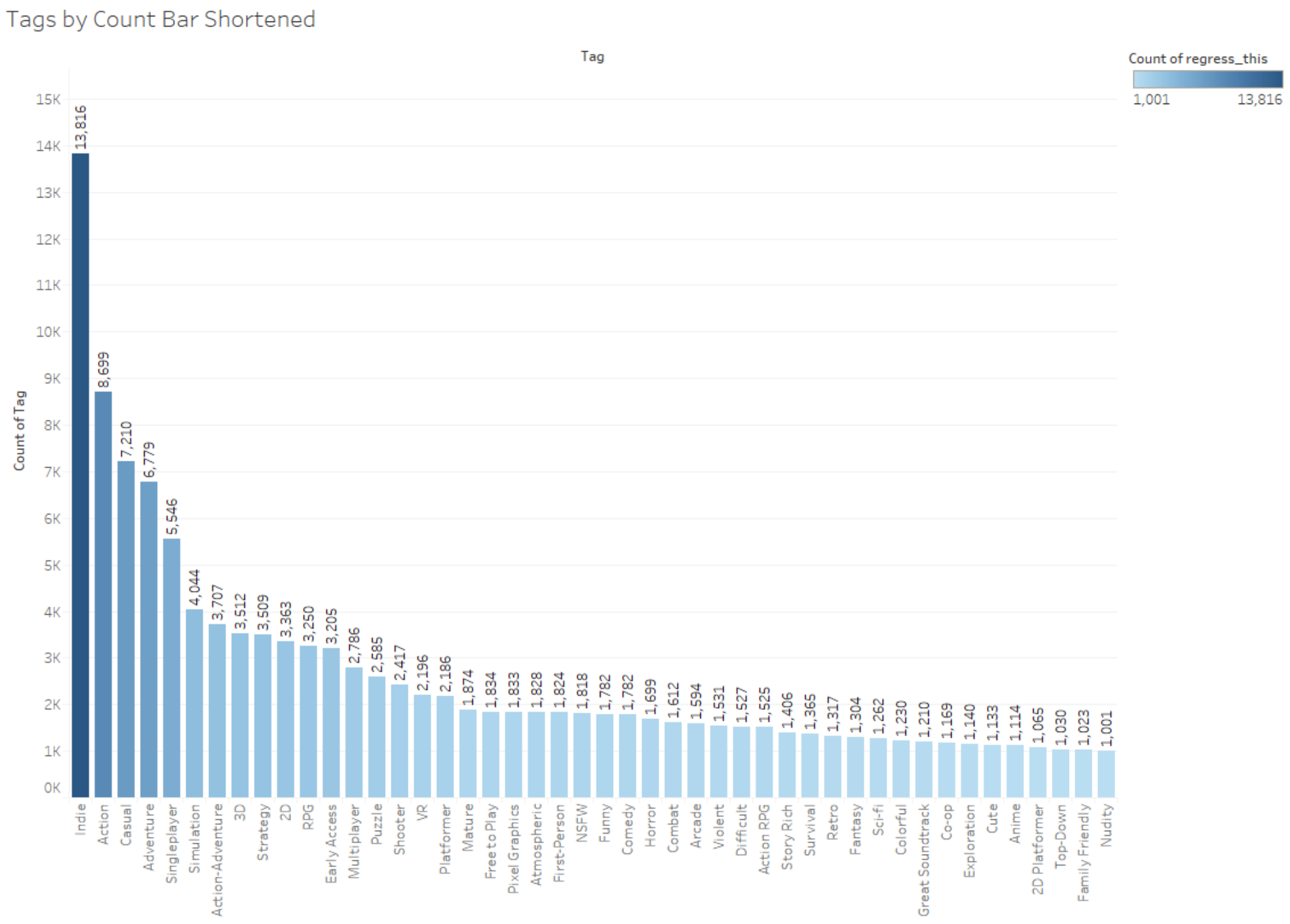

*Figure 5. Truncated bar chart showing count of each tag in dataset*

## Appendix B: G Square Conditional Independence Tests Results

| conditional independence | G Square p-value | Independent? |
|---|---|---|
| $2D \perp\!\!\!\perp twitter$ | 0 | No |
| $3D \perp\!\!\!\perp twitter$ | 0.014 | No |
| $Action \perp\!\!\!\perp twitter$ | 0.713 | yes |
| $Action\text{-}Adventure \perp\!\!\!\perp twitter$ | 0.016 | No |
| $Adventure \perp\!\!\!\perp twitter$ | 0 | No |
| $Casual \perp\!\!\!\perp twitter$ | 0.016 | No |
| $EarlyAccess \perp\!\!\!\perp twitter$ | 0 | No |
| $Indie \perp\!\!\!\perp twitter$ | 0.148 | yes |
| $Multiplayer \perp\!\!\!\perp twitter$ | 0 | No |
| $Puzzle \perp\!\!\!\perp twitter$ | 0 | No |
| $RPG \perp\!\!\!\perp twitter$ | 0.235 | yes |
| $Shooter \perp\!\!\!\perp twitter$ | 0.001 | No |
| $Simulation \perp\!\!\!\perp twitter$ | 0.008 | No |
| $Singleplayer \perp\!\!\!\perp twitter$ | 0.008 | No |
| $Strategy \perp\!\!\!\perp twitter$ | 0.002 | No |
| $VR \perp\!\!\!\perp twitter$ | 0 | No |

*Table 4. Results of the G Square conditional independence test conducted in casualfusion.net, showing mixed results for [tag] $\perp\!\!\!\perp$ twitter*